# Synchronization Implies Seizure or Seizure Implies Synchronization?

Kaushik Majumdar • Pradeep D. Prasad • Shailesh Verma

*Systems Science and Informatics Unit, Indian Statistical Institute, 8th Mile, Mysore Road, Bangalore 560059*

Ph: +91-80-2848-3002; Fax: +91-80-2848-4265; E-mail: kmajumdar@isibang.ac.in (KM), pradeepd.prasad@gmail.com (PDP), shaileshvermadrtc@gmail.com (SV)

**Abstract**   Epileptic seizures are considered as abnormally hypersynchronous neuronal activities of the brain. The question is "Do hypersynchronous neuronal activities in a brain region lead to seizure or the hypersynchronous activities take place due to the progression of the seizure?" We have examined the ECoG signals of 21 epileptic patients consisting of 87 focal-onset seizures by three different measures namely, phase synchronization, amplitude correlation and simultaneous occurrence of peaks and troughs. Each of the measures indicates that for a majority of the focal-onset seizures, synchronization or correlation or simultaneity occurs towards the end of the seizure or even after the offset rather than at the onset or in the beginning or during the progression of the seizure. We also have outlined how extracellular acidosis caused due to the seizure in the focal zone can induce synchrony in the seizure generating network. This implies synchronization is an effect rather than the cause of a significant number of pharmacologically intractable focal-onset seizures. Since all the seizures that we have tested belong to the pharmacologically intractable class, their termination through more coherent neuronal activities may lead to new and effective ways of discovery and testing of drugs.

**Keywords**   Electrocorticography (ECoG), Focal epilepsy, Hilbert phase synchronization, Amplitude correlation, Peaks and troughs detection, Seizure termination, Extracellular acidosis.

## Introduction

The words seizure and synchronization almost go hand in hand. The International League Against Epilepsy (ILAE) and the International Bureau of Epilepsy (IBE) have come to a consensus definition for the term *epileptic seizure*. According to this definition an epileptic seizure is a transient occurrence of signs and/or symptoms due to abnormal excessive or synchronous neuronal activity in the brain [1]. Way back in 1954 Penfield and Jasper [2] were the first to postulate hypersynchronization as the possible cause for seizure [3]. Since then the presumption that the seizures are a manifestation of excessive abnormal synchronization of neuronal firing, either locally or globally in the brain, has become pervasive [4].

Synchronization in seizures has however created myriad different notions among researchers. Some observed decreased synchronization before the onset of seizure [5, 6]. Others observed decreased synchronization [7] and decreased correlation [8] among the ECoG focal channels during the seizure, which actually goes against the hypersynchrony doctrine. More recently single cell recording in the focal onset zone and beyond has revealed a complex pattern of evolution of synchrony during the seizure [9]. In the beginning of the seizure neuronal firing remains more heterogeneous, but gradually becomes synchronous towards the end. Enhanced synchronization towards the end of the seizure rather than in the beginning has also been observed through the study of amplitude correlation in the ECoG data of focal epileptic patients [10]. One important aspect to note is the various ways the notion of synchrony have been interpreted in these studies. Phase synchrony has been defined by means of phase coherence on the instantaneous phases obtained by Hilbert transformation [5] and by wavelet phase coherence [6]. Synchrony has been tested in terms of cross correlation among the time series data, mutual information and instantaneous phase difference in [7]. In [9] neuronal spike trains have been compared for occurrence of spikes (variance divided by mean of the spikes) to determine the synchrony. Averaged coherence of ECoG signals showed an increasing trend from onset to the offset of temporal lobe seizures [11]. Although different measures produced different results on different types of depth recordings, they do have certain common trends, which indicate complicated patterns of synchronization from before the onset up to the offset of the seizures.

Epileptic seizures are caused by excessive discharges of a collection of neurons either in the neocortex or in the subcortical structures. Interestingly, synchronous firing among single neurons within the epileptic cortex was not that apparent [12]. Recent and more elaborate single cell studies in the epileptogenic zone have revealed a heterogeneous neuronal firing pattern from prior to the onset till the offset of seizure [9], where synchronization (as homogeneous firing of neuronal population) has been observed to be a progressive phenomenon from onset to offset. This is consistent with the finding by MEG studies of generalized seizures that seizures proceed by a recruitment of neighboring neuronal networks [13]. Seizure is generated by excitatory pyramidal neurons, which sit in the layer 3 and layer 5 of the cortex. It has been found in genetically modified rat model that generalized absence seizures are initiated by layer 5 pyramidal neurons [14, 15]. If that is the case for human focal epilepsy also, subdural ECoG is not supposed to capture the seizure generating cellular spike trains. However it has been shown by ECoG studies of epileptic patients during status epilepticus that increased synchronization is actually indicative of termination of seizure [16]. It has even been observed that anticonvulsant drugs induced synchronization during termination of seizure [16].

Almost all the epileptic seizures share a remarkable property – they terminate on their own. A lot more studies have been undertaken on the generation of seizures than on the termination. Cellular mechanisms of termination have started to be investigated more recently and are open avenues of research [17]. But a deeper insight into the termination process may help developing new and more effective anticonvulsant drugs [18]. This is particularly important for the pharmacologically intractable epilepsies, a portion of which are only curable by surgical intervention and therefore have to undergo pre-surgical evaluation through the subdural grid ECoG recording. This adds significance to the new found relation between termination of seizure and multichannel synchronization in the ECoG recording. Causes of seizure termination are however quite diverse [18, 19], all of which may not be related to synchronization. One line of evidence suggests seizure can turn an epileptogenic zone from mild alkaline to mild acidic [19]. Acidity is known to suppress the activation of $Na^+/Ca^{++}$ channels in the membrane [20] and thereby suppresses hyperexcitability of the excitatory neurons. Acidity also modulates $GABA_A$ receptors [20], thereby contributing to inhibition. Acidity also facilitates activation of inhibitory neurons [21]. It is well established by modeling studies that combination of excitatory and inhibitory neurons in a network is necessary to induce synchronized firing patterns [22, 23].

With the outcome of three different and independent measures of synchronization on the ECoG data set [24] of 87 focal-onset seizures of 21 patients we have shown that in a majority of the cases it is the seizure which progressively imposes synchronization starting from onset to offset and beyond. This is compatible with the recent experimental finding reported in ref. [9]. Next we have investigated if synchronization has any role in seizure termination. It is already known that synchronization helps terminate status epilepticus [16]. In terms of extracellular acidosis induced by seizure in the epileptogenic zone we have proposed a seizure termination dynamics in which seizure implies synchronization and synchronization leads to termination. This might be the case for many pharmacologically intractable focal-onset seizures. Study of synchronization using different measures alongside monitoring the pH level of the epileptogenic zone may give ways to develop and testing novel drugs.

**Methods**

*Hilbert phase synchronization*

In this paper the phase of a time domain signal will imply the instantaneous phase determined by Hilbert Transform [25] and the corresponding synchronization is referred to as *Hilbert phase synchronization*. This metric is evaluated between two ECoG channels as phase locking between two periodic oscillators [25, 26]

$$|\varphi_{n,m}(t)| < c,$$

$$\varphi_{n,m}(t) = n\varphi_1(t) - m\varphi_2(t), \quad (1)$$

where $c$ is a constant, $\varphi_1, \varphi_2$ are phases of the ECoG from the first and the second channel respectively. Equation (1) describes $n:m$ phase locking between two signals. In the current work, we consider the $n = m = 1$ case only. Let $s(t)$ be any time domain signal. Let us define

$$\psi(t) = s(t) + j \cdot \hat{s}(t) = A(t) \cdot \exp(j \cdot \varphi(t)) \quad (2)$$

where $\hat{s}(t)$ is the Hilbert transform of the signal $\hat{s}(t)$, $j = \sqrt{-1}$, $A(t)$ is the *envelope* (*instantaneous amplitude*) of $\psi(t)$ and $\varphi(t)$ is the (instantaneous) phase of $\psi(t)$. We define $\varphi(t)$ and $A(t)$ to be the *instantaneous phase* and *envelope* of $s(t)$ respectively.

$$\hat{s}(t) = \frac{1}{\pi} \int_{-\infty}^{\infty} \frac{s(t)}{t - \tau} d\tau \quad (3)$$

In equation (3) only the Cauchy principle value has been considered. For the interpretation of $A(t)$ and $\varphi(t)$ to be meaningful, the signal $s(t)$ must have to be in a narrow band [25]. The representation in eq. (2) allows one to remove the influence of amplitude from the phase synchronization measure between two signals and thus allowing an independent measurement of phase locking between them.

It has been shown in [27] that when two signals are in phase synchrony $\varphi_{n,m}(t)$ cycles through a set of 'preferred' values. If there is no synchrony then $\varphi_{n,m}$ will exhibit a large number of values. The distribution of phase differences between two signals that are not in phase synchrony would thus resemble a uniform distribution.

Thus, quantifying the deviation of the obtained distribution of phase differences from a uniform distribution would provide a measure of phase synchronization. The Shannon entropy can be used to quantify such a deviation [27]. Since the unique value of phase difference can take on values in $[0, 2\pi]$, we divide the interval into $N$ number of phase bins. Let us denote the Shannon entropy of the phase differences across all the bins by $S$. Then

$$S = -\sum_{i=1}^{N} p_i \ln(p_i) \qquad (4)$$

where $p_i$ is the probability of the phase difference being in the $i^{th}$ bin. By computing $S_{max} = \ln(N)$, the entropy corresponding to the uniform distribution, we can compute a normalized measure of phase synchronization ($\gamma$)

$$\gamma = 1 - \frac{S}{S_{max}}. \qquad (5)$$

If $\gamma = 1$ signals are in perfect phase synchrony and if $\gamma = 0$ the signals are in perfect phase asynchrony.

The above method can be used to determine the phase synchronization between a pair of signals for any duration of time by using a sliding window. The signals are first band-pass filtered in order to get a narrow band with respect to which the instantaneous Hilbert phase has to be determined. Consider a window of size $T$ seconds starting at time $t = t_0$, i.e., segments of the signal in the interval $[t_0, t_0 + T]$. We compute a value of $\gamma$ corresponding $t = t_0$. We next shift the window to time $t = t + \delta t$ and compute $\gamma$. This procedure is repeated until the synchronization has been computed over the requisite time interval of interest to obtain a synchronization profile that varies over time.

Note that the integral in eq. (3) is a convolution and therefore it can be computed in a time efficient manner by using the Fast Fourier Transform ($F$).

$$\hat{s}(t) = F^{-1}(-j\,\text{sgn}(w)F(s)) \qquad (6)$$

where $\text{sgn}(\ )$ is the signum function. The phase of the signal is then extracted using eq. (2) for all points in the interval $[t_0, t_0+T]$. Then the interval $[0, 2\pi]$ is divided into $N$ bins and the distribution of the phase differences are computed. Upon computing the entropy of this distribution using eq. (4), $\gamma$ can be evaluated by eq. (5).

After computing $\gamma$, we test the null hypothesis of no synchronization, i.e., the synchronization we observe is not significant or $H_0 : \gamma = 0$. The decision rule we use to test this hypothesis is : Reject $H_0$ if $\gamma > \gamma_0$. For a given p-value, $\gamma_0$ is chosen as the $100(1-p)$ percentile of the distribution of values of $\gamma$ obtained by evaluating the phase synchronization between a large number of pairs of independent shifted-time surrogate signals. It has been shown that this method of obtaining $\gamma_0$ is equivalent to using white noise signals in place of shifted-time surrogates [28].

It should be noted that it is possible to use the wavelet transform to extract phase and amplitude information from the signals [28]. However, it has been shown that the results obtained by using wavelet transform are virtually indistinguishable from those obtained by the Hilbert transform [29].

*Amplitude correlation*

If there are $r$ number of channels, then $r \times r$ cross-correlation matrix has to be formed. The matrix is calculated for cross-correlation over a window with $m$ time points. Then $r$ eigen values of the matrix are calculated and sorted in descending order. Then the window is slid (usually continuously) and the process is repeated. The temporal plot of the highest eigen value is generated by the highest eigen values at all time points. If the highest eigen value plot is increasing with respect to time, it is said that the overall amplitude correlation is growing up. If it is decreasing, the overall correlation is also decreasing. For the detail see [10].

*Multi-channel peaks and troughs detection*

The peaks and troughs in a digital signal $s$ will appear as in Fig. 1. Let the coordinate of B be $(m, s(m))$, that of A be $(m-1, s(m-1))$ and of C be $(m+1, s(m+1))$. Clearly, $s(m) - s(m-1) > 0$, $s(m+1) - s(m) < 0$. Note that $s''(m) = s(m+1) - s(m) - (s(m) - s(m-1)) < 0$. So $P(m-) = s''(m)(s(m) - s(m-1)) < 0$ and $P(m+) = s''(m)(s(m+1) - s(m)) > 0$. Therefore in case of digital signal a peak at $m$ can be

identified by the property

$$P(m-) < 0 \ \& \ P(m+) > 0 \ \& \ s''(m) < 0. \tag{7}$$

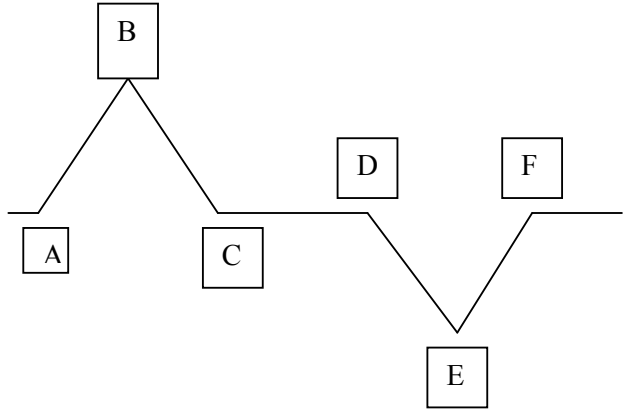

Fig. 1. A peak (B) and a trough (E) in a digital signal.

Similarly, the trough at E in Fig. 1 with coordinate $(k, s(k))$ can be identified by the property

$$P(k-) < 0 \ \& \ P(k+) > 0 \ \& \ s''(k) > 0. \tag{8}$$

Note that a difference operation works as a high-pass filter. The operator $P$ will therefore work as a high-pass filter leading to allowing the high frequency noise to pass through. To avoid the former a low-pass filter should be used before applying $P$. Regarding the latter, note that we are not using the signal morphology after being operated by $P$. We are only using sign of $P$ and $s''$ to identify a peak (eq. (7)) or a trough (eq. (8)), which remains true for peaks and troughs irrespective of their shape, amplitude or any other morphological properties. This is one major advantage of the proposed algorithm over any other morphology based algorithm. Therefore $P$ acting as a high-pass filter is not affecting the detection performance.

For a two-channel simultaneous (with zero time lag) peak or trough detection let $U_r$ and $V_r$ are two $r$ dimensional vectors (time domain signals) with only binary entries, i.e., $0$, for no peak or no trough, and 1 for a peak or a trough. By a *Boolean vector product* of $U_r$ and $V_r$ we mean the $r$ dimensional vector $W_r$, where the $i$th entry of $W_r$ is the Boolean product of the $i$th entries of $U_r$ and $V_r$ for all $i \in \{1,.....,r\}$. Let us express this as $U_r |*| V_r = W_r$, where $|*|$ denotes the Boolean vector product. It is easy to see for any $p$ number of channels $U_r^1(t).U_r^2(t).U_r^3(t)..........U_r^p(t) = 1$ if there is a peak (trough) at $t$ in $U_r^i(t)$ for every $i$.

**Patients and data**

TABLE I

| Patient | Gender | Age | Seizure Type | H/NC | Electrode | Origin | # seizures |
|---|---|---|---|---|---|---|---|
| 1 | F | 15 | SP,CP | NC | g,s | Frontal | 4 |
| 2 | M | 38 | SP,CP,GTC | H | D | Temporal | 3 |
| 3 | M | 14 | SP,CP | NC | g,s | Frontal | 5 |
| 4 | F | 26 | SP,CP,GTC | H | D,g,s | Temporal | 5 |
| 5 | F | 16 | SP,CP,GTC | NC | g,s | Frontal | 5 |
| 6 | F | 31 | CP,GTC | H | D,g,s | Temporo/Occipital | 3 |
| 7 | F | 42 | SP,CP,GTC | H | D | Temporal | 3 |
| 8 | F | 32 | SP,CP | NC | g,s | Frontal | 2 |
| 9 | M | 44 | CP,GTC | NC | g,s | Temporo/Occipital | 5 |
| 10 | M | 47 | SP,CP,GTC | H | D | Temporal | 5 |
| 11 | F | 10 | SP,CP,GTC | NC | g,s | Parietal | 4 |
| 12 | F | 42 | SP,CP,GTC | H | D,g,s | Temporal | 4 |
| 13 | F | 22 | SP,CP,GTC | H | d,s | Temporo/Occipital | 2 |
| 14 | F | 41 | CP,GTC | H and NC | d,s | Fronto/Temporal | 4 |
| 15 | M | 31 | SP,CP,GTC | H and NC | d,s | Temporal | 4 |
| 16 | F | 50 | SP,CP,GTC | H | d,s | Temporal | 5 |
| 17 | M | 28 | SP,CP,GTC | NC | S | Temporal | 5 |
| 18 | F | 25 | SP,CP | NC | S | Frontal | 5 |
| 19 | F | 28 | SP,CP,GTC | NC | S | Frontal | 4 |
| 20 | M | 33 | SP,CP,GTC | NC | D,g,s | Temporo/Parietal | 5 |
| 21 | M | 13 | SP,CP | NC | g,s | Temporal | 5 |

H = hippocampal, NC = neocortical, SP = simple partial, CP = complex partial and GTC = generalized tonic clonic.

ECoG data of 21 epileptic patients containing 87 focal-onset seizures have been obtained from the Freiburg Seizure Prediction Project [24]. One hour recording containing preictal, ictal and postictal ECoG of one hour duration in each of the 87 cases is available. The ECoG data were acquired using Neurofile NT digital video EEG system (It-med, Usingen,

Germany) with 128 channels, 256 Hz sampling rate, and a 16 bit analog to digital converter. In all cases the ECoG from only six sites have been analyzed. Three of them from the focal areas and the other three from out side the focal areas (this is the configuration with which the data was available to us). See Table 1 for the patient details, seizure types and their focal locations. For each patient there are two to five hours of ictal data (actually preictal + ictal + postictal) and twenty four to twenty six hours of interictal data (except patient 2, for whom no interictal data is available).

## Results

*Hilbert phase synchronization*

Hilbert phase synchronization has been studied on all the 87 seizures at the alpha, beta and lower gamma (30 – 40 Hz) range. We have chosen only the lower gamma range instead of the whole gamma because (i) it is in narrow band, for which phase synchronization measure is more appropriate [25] and (ii) 40 Hz gamma rhythm is prototypical of the gamma band (30 – 100 Hz) oscillation [30].

Let S denote the statistical significance value, which is equal to the maximum Hilbert phase synchronization value in the 95% of the 100 pairs of shifted surrogate signals. The entire seizure duration has been divided into two equal parts. Maximum synchronization during first half of the seizure period is denoted as M1 and maximum synchronization during second half of the seizure duration is denoted as M2. If M1 is lesser than the significance level S then the threshold is set to S. If M1 is greater than S then the threshold is set to M1. The criteria to satisfy is — M2 should be greater than or equal to 1.25 times the threshold and minimum value of 0.1 (If M2 < 0.1, we ignore it altogether). The outcome of the classification has been presented in the Table II below.

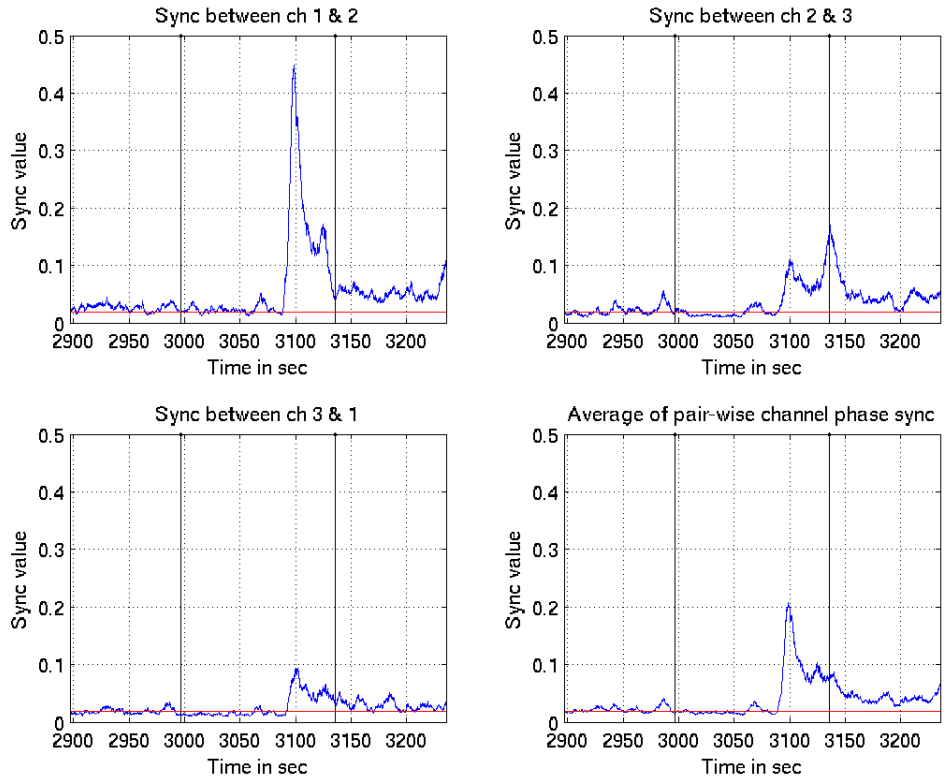

Fig. 1. Pairwise and average (average of all pairs) phase synchronization in the lower gamma band among three focal channels during seizure of patient 2 in the 21st hour of recording. The vertical lines denote seizure onset and offset points. The bottom horizontal line denotes the statistical significance value.

Table II

| A total of 87 seizures recorded from the focal ECoG of a total of 21 focal epileptic patients has been tested | Ch. 1 & 2 | Ch. 2 & 3 | Ch. 3 & 1 | Average of pairwise synch |
|---|---|---|---|---|
| Total number of seizures showing the criterion out of a total of 87 seizures | 45 (52%) | 34 (39%) | 36 (41%) | 39 (45%) |
| Number of patients showing the criterion at least once out of a total of 21 patients | 17 (81%) | 17 (81%) | 17 (81%) | 18 (86%) |
| Fraction of seizures only from the patients showing the criterion at least once | 45/73 (62%) | 34/72 (47%) | 36/72 (50%) | 39/75 (52%) |

Here the word 'criterion' means the enhancement of phase synchronization during the second half of seizure compared to the first half.

From the Table II it is clear that channels 1 and 2 are the most predominant focal channels across the patient population in the sense that most number of seizures were generated in their location (this will become even more clear in Table III). It has been observed in ref [10] that for secondary generalized seizures all seizures showed enhanced amplitude correlation towards the end of the seizure. Our data shows that for focal-onset seizure patients whose Hilbert phase synchronization goes up towards the end of seizure at least for once, 45% of the total number of the seizure occurrences in them have average pairwise higher phase synchronization in the second half of the seizure duration than in the first.

*Amplitude correlation*

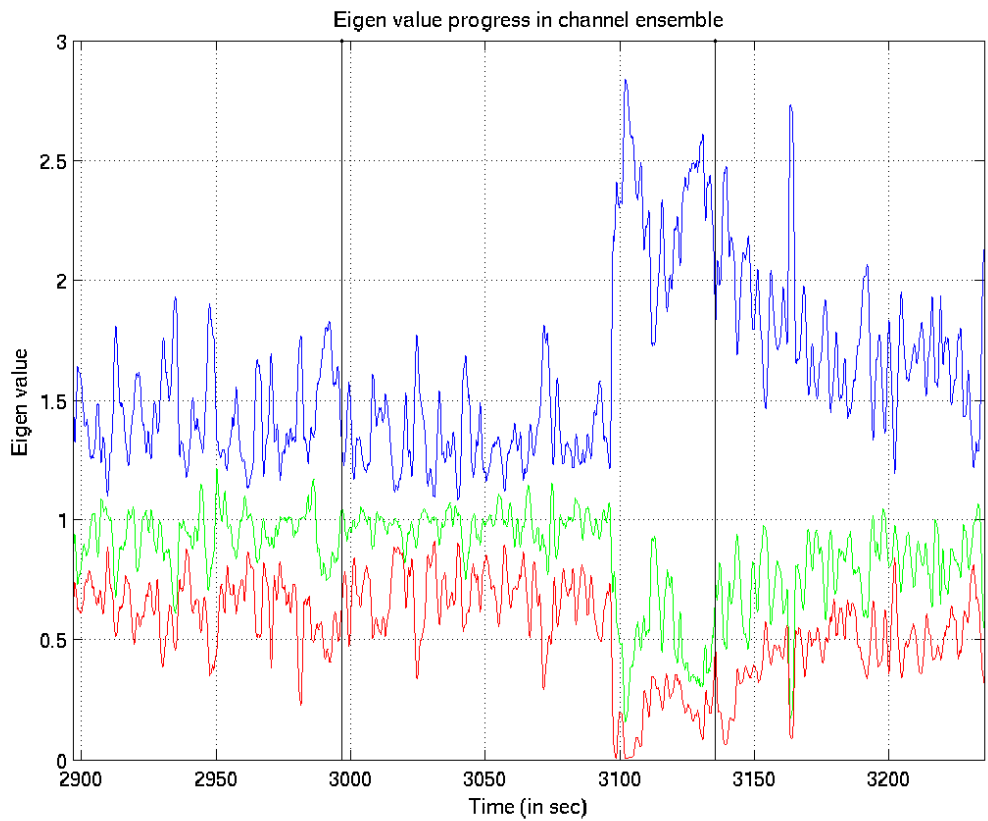

Fig. 2. Three eigen values of the amplitude correlation matrix during the 21st hour of recording of patient 2. Vertical lines indicate seizure onset and offset points.

We have tested amplitude correlation at different frequency bands in 0 – 40 Hz range. In all the bands the highest eigen value remained higher in the second half of the duration of the seizure than the first in majority of the cases (Fig. 2). But for the sake of comparison here we will present results of only lower gamma band. In the next subsection we will see coherence is more probable in seizures at the higher frequency bands than in the lower. However amplitude correlation has been studied in [10] across all the available bands, i.e., without any prior band pass filtering, possibly apart from the one already built in the acquisition system. Also complex patterns of amplitude correlation and decorrelation have been observed in the limbic networks prior to onset of temporal lobe seizures [31]. During seizure excessive correlation between cortical and subcortical structures may lead to loss of consciousness [32].

The preprocessing methods and bandwidth selection remain the same as in the phase synchronization study. A 2 second long window with continuous shifting has been used. The results have been smoothed with a moving average filter of 1 second duration for a less jittery visualization of the trend (Fig. 2). The criterion for identifying a seizure as having high amplitude correlation during seizure offset is as following. During the first half of seizure, the maximum value that the highest eigen value reaches is denoted as M1 and during second half of seizure the maximum value is denoted as M2. The criterion to satisfy is that M2 should be at least 1.1 times M1.

Table III

| A total of 87 seizures recorded from the focal ECoG of a total of 21 focal epileptic patients have been tested | Amplitude correlation |
|---|---|
| Total number of seizures showing the criterion | 51 (59%) |
| Number of patients showing the criterion at least once | 19 (90%) |
| Total number of seizures only from the patients showing the criterion at least once | 80 (92%) |

Here the word 'criterion' means the enhancement of amplitude correlation during the second half of seizure compared to the first half.

Expectedly amplitude correlation is showing a greater trend than phase synchronization to be higher towards the end of the seizure rather than in the beginning. This is because phase synchronization is a more sensitive measure than amplitude correlation. However they offer two different kinds of insights into the underlying dynamical systems generating the time series. We have listed in Table IV the seizure hours where phase synchronization is more towards the end than in the beginning, but amplitude correlation is not more towards the end compared to the beginning and vice versa.

Frontal lobe seizures are known to be quite complex in nature. From Table IV it is observed that patient 19 with frontal lobe seizure (Table I) has an hour (69th hour) when amplitude correlation is more towards the seizure offset than after the onset and has another hour (15th hour) when phase synchronization is more towards the offset than after the onset. In 69th hour phase synchronization is less towards the seizure offset than after the onset and in 15th hour amplitude correlation is less towards the seizure offset than after the onset. Patient 19 is the only one out of 21 patients having a seizure with higher amplitude correlation but not higher phase synchronization towards the end and also higher phase synchronization but not higher amplitude correlation towards the end.

TABLE IV

PHASE SYNCHRONIZATION VERSUS AMPLITUDE CORRELATION

| Patient number | Seizure hour with enhanced phase synchronization towards the offset | Seizure hour with enhanced amplitude correlation towards the offset |
|---|---|---|
| 2 | $15^{th}$ | - |
| 3 | - | $176^{th}$ |
| 5 | - | $13^{th}, 26^{th}, 33^{rd}$ |
| 9 | - | $26^{th}, 38^{th}$ |
| 10 | $200^{th}$ | - |
| 11 | - | $5^{th}$ |
| 15 | - | $41^{st}$ |
| 16 | - | $8^{th}$ |
| 17 | - | $99^{th}$ |
| 18 | - | $14^{th}$ |
| 19 | $15^{th}$ | $69^{th}$ |
| 20 | - | $13^{th}, 55^{th}, 82^{nd}$ |

Seizure hour means one hour recording of pre-ictal, ictal and post-ictal ECoG during the specified numbered hour.

*Simultaneous peaks and troughs*

Peaks and troughs are important characteristics of electrophysiological signals. Although ECoG recordings cannot directly measure the synchronization of action potentials among assemblies of neurons, they may demonstrate event-related interactions in different frequency bands in macroscopic local field potentials (LFP) generated by different large-scale populations of neurons engaged by a specific occurrence [33], such as a seizure. Unlike scalp EEG, ECoG is largely noise free, it may still have artifacts due to movement, muscle contraction, electrode scratching etc. Peaks and troughs in ECoG may be due to artifacts also. However when there is a peak due to genuine neuronal firing it may signify a simultaneous depolarization of a large number of neurons within a neighborhood of the point on the surface of the cortex where the subdural electrode has been implanted. Similarly a trough in an ECoG signal may signify hyperpolarization of a large number of neurons within the neighborhood of the point on the surface of the cortex where the subdural electrode has been implanted. Simultaneous peaks and troughs across the focal ECoG channels are therefore indicative of simultaneous firing of neurons in the seizure focal regions (Fig. 3). However exceptions to this have recently been reported [34]. Fig. 3 clearly shows that the number of simultaneous peaks and troughs across the focal channels after the seizure offset is much greater within the same time span as the seizure duration than the number of simultaneous peaks and troughs across the same channels during the seizure. Actually in the 0 – 40 Hz range in 82.28% of the cases simultaneous peaks are more across focal channels after the seizure offset than during the seizure and in 77.22% of the cases troughs are more. Interestingly the trend for non-focal channels is weaker. Across the non-focal channels simultaneous occurrence of peaks is more after the seizure for 55.7% of the cases and for the troughs it is 49.37%. That lesser simultaneous events should occur in the non-focal channels compared to the focal has been reported in [34, 35]. It has been reported in [8] that amplitude correlation among ECoG channels during frontal and temporal lobe epilepsies goes down compared to before the onset and after the offset. Correlation and decorrelation at different scales before, during and after the seizure have been reported in [36].

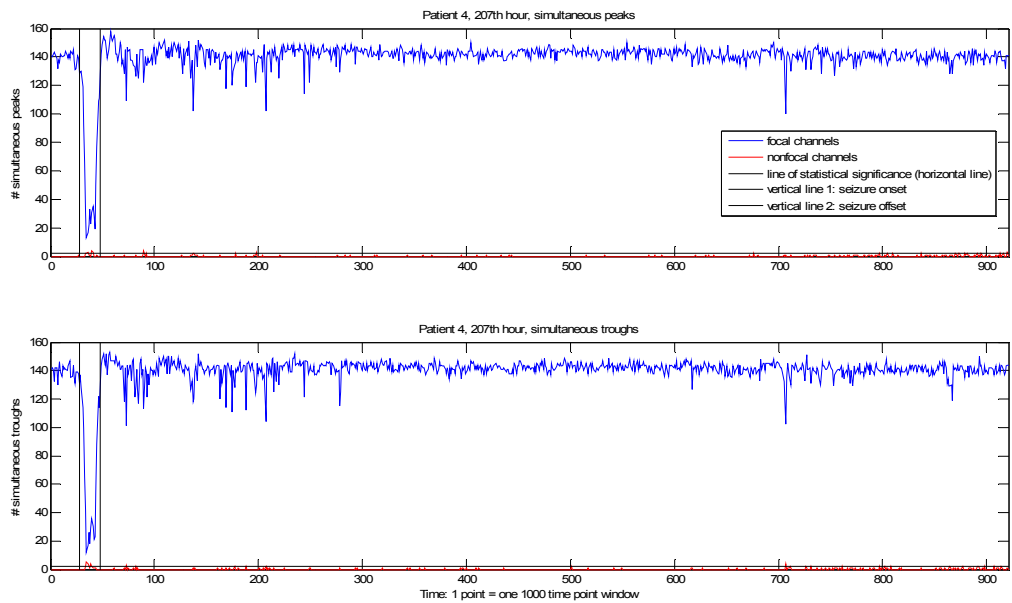

Fig. 3. Simultaneous peaks (top panel) and troughs (bottom panel) of the patient 4, during 207th hour of recording for 3 focal (in blue) and 3 non-focal (in red) channels. The pair of vertical lines denote seizure onset and offset. The bottom horizontal line denotes statistical significance value. Here 1 time point = 1000 sample point window. In this case the signals had been preprocessed with Gaussian low-pass filter with cut off frequency 40 Hz and stop-band attenuation 40 dB.

Table V

| Band | Peaks | Troughs |
|---|---|---|
| Delta | 20% | 27.5% |
| Theta | 51.12% | 51.12% |
| Alpha | 67.5% | 65% |
| Beta | 75% | 71.25% |
| Lower Gamma | 80% | 81.25% |

Peaks = Percentage of the seizures in which the number of peaks is more after the offset than during the seizure. Troughs = Percentage of the seizures in which the number of troughs is more after the offset than during the seizure.

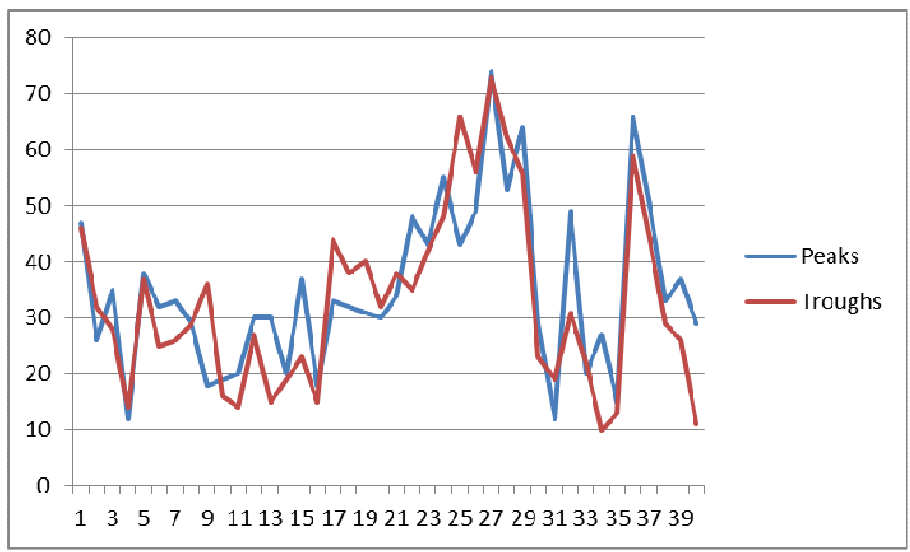

Fig. 4. Number of simultaneous peaks (blue) and troughs (red) across focal channels in delta band in every hour for all hours across all the interictal and ictal periods of patient 17. Ordinate gives the number of simultaneous peaks and troughs in an hour and abscissa indicates the hours, with the hours, in which a seizure occurred, towards the end.

Next we compared the simultaneous occurrence of peaks and troughs across the focal channels during and after the seizure for delta, theta, alpha, beta and lower gamma bands. The trend goes up as the frequency goes up (Table V). Note that in this study we preprocessed the signals with equiripple band-pass filter. Upon checking with both Gaussian and equiripple filters it appeared that the number of peaks and troughs varied, but the outcome of our study in 0 – 40 Hz range remained the same. The interesting thing to notice is the increasing number of simultaneous peaks and troughs after the seizure than during the seizure occur at progressively higher frequencies. Is the focal-onset seizure synchronization likely to occur more in the higher frequency range than in the lower frequency range? The answer seems to be in affirmative because high-frequency epileptiform oscillations (HFEO, > 60 Hz) are involved in neocortical focal-onset seizure generation [37] and very high frequency (HFO, > 100 Hz) oscillations are involved in the human focal-onset seizure generation [38]. Also low voltage, higher than 25 Hz signal frequency has been observed to predominate the ECoG signals during focal seizures [8].

Since every depolarization is followed by a hyperpolarization in a single cell firing it is only natural that in LFP and ECoG recordings the number of peaks and troughs will closely match each other. We have observed that this is true in frequency bands from alpha through lower gamma, except in delta (Fig. 4). In fact in the lower gamma range they match almost perfectly (Fig. 5).

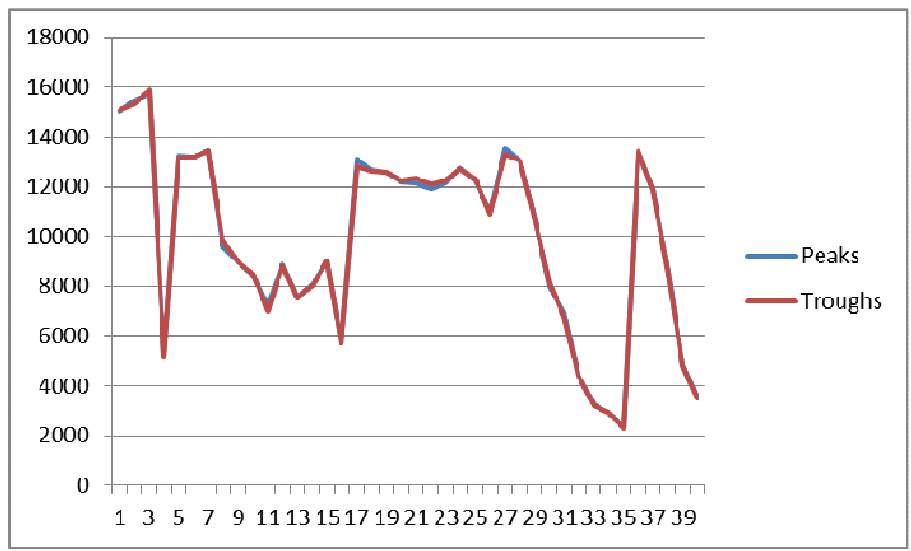

Fig. 5. Same as in Fig. 4, but for the lower gamma range.

**Termination dynamics**

It is a remarkable fact that almost all seizures terminate on their own. For the focal-onset seizures of hippocampal origin it has been hypothesized that intense firing by pyramidal cells during a seizure results in reduced concentration of $K^+$ inside the cell, which signals the astrocytes to release $K^+$. This induces a depolarization block of spike generation in neurons, resulting in postictal depression of the electrical signals [39]. As the postictal depression sets in within the hyperexcited seizure generating network, the network may undergo through synchronous behavior.

Actually seizure termination involves multiple mechanisms. One line of evidence suggests that extracellular acidity in the focal region may terminate seizure [19]. Seizure can reduce brain pH from ~ 7.35 to ~ 6.8 through lactic acid production, $CO_2$ accumulation and other mechanisms [19]. Acidity inhibits the transmission of the $Na^+$ and $Ca^{++}$ ions through cell membrane [20] and thereby reducing the hyperexcitability of the pyramidal cells. Acidosis inhibits the NMDA receptor and abets the $GABA_A$ receptor [20], thereby suppressing excitation and abetting inhibition. At the same time it may excite inhibitory interneurons, because they have larger $H^+$ gated current, which is facilitated by extracellular acidity [21]. From computational modeling it is known that phase locking in a network is maintained by interconnected excitatory and inhibitory neurons in the network [22, 23]. Here we hypothesize – as seizure progresses extracellular acidity increases leading to diminished firing of excitatory pyramidal neurons and enhanced firing of inhibitory interneurons. This induces greater synchronization in the seizure focal network towards the end of seizure as evident from computational modeling [22, 23]. We can encapsulate our hypothesis in the following diagram (Fig. 6).

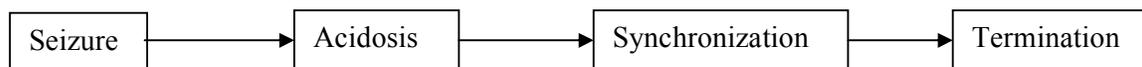

Fig. 6. Seizure termination through synchronization.

**Conclusion**

Epileptic seizure has traditionally been perceived as a manifestation of an abnormally high neuronal synchrony. But this is a rather general notion. There are at least two important questions remain to be answered in order to make the notion more precise. (1) There are many forms of synchronization [40]. Which synchronizations go high during the seizure? (2) What is the spatio-temporal pattern of evolution of synchronization during the seizure? In this paper we have addressed both the questions, albeit partially. We have shown in the seizure focal region Hilbert phase synchronization and amplitude correlation go lower to higher in the majority of the cases as the seizure progresses from the onset to the offset.

We have also introduced a new measure, which is not a measure of synchronization per se, but captures many of the essences of phase synchronization and amplitude correlation. The new measure is simultaneous occurrence (with zero time lag) of peaks and troughs across the focal channels. When two peaks occur simultaneously in two time series data and their shapes are not drastically different they are likely to have a high degree of phase synchronization and amplitude correlation. Same is true for simultaneous troughs. But in between two simultaneous peaks the two time series may have very low degree of phase synchronization and amplitude correlation. Nevertheless peaks and troughs carry the most significant information in electrophysiological signals, particularly during the seizure. Simultaneous occurrence of peaks and troughs is a good measure of synchronous bursts (consisting of trains of depolarization, hyperpolarization and again depolarization constituting a peak and a trough) of neuronal ensemble, which is a hallmark of epileptic seizure. We have shown that in the entire seizure focal region this simultaneous bursts of neuronal ensemble occur more after the offset than during the seizure. This raises the question, "Is synchronization more important for starting the seizure or terminating the seizure?" This question has been bolstered by a recent experimental finding that firing of neurons in the focal region becomes more heterogeneous to more homogeneous with the progression of the seizure from before the onset till the offset [9]. There are many different ways seizures can start and terminate. Putting together some experimental findings we have been able to hypothesize a termination dynamics for focal-onset seizures (Fig. 6). According to this dynamics seizure implies synchronization and synchronization leads to termination (this can happen even without extra cellular acidosis, for example, through ephaptic coupling, see ref. [41]). This may elucidate us about the mystery, "Why almost all seizures terminate on their own?" The dynamics in Fig. 6 may help in developing and evaluating the effectiveness of anticonvulsant drugs.

In our study simultaneous peaks and troughs has turned out to be a more definitive measure compared to phase synchronization and amplitude correlation, in the sense that when these two measures did not show any specific trend peaks and troughs measure was still able to identify a trend. Number of peaks and troughs varies according to the filter chosen for preprocessing, but the trend always remains the same. It is to be noted that thalamus and corticothalamic loops play an important role in human temporal lobe seizures [42, 43]. It was observed that the average correlation among focal ECoG channels were more at the offset than at the onset for most of the seizures [42]. This indicates thalamus may play a role in having more number of peaks and troughs after the temporal lobe seizure offset than during the seizure.

**Acknowledgement**

This work has been supported by an internal research grant of the Indian Statistical Institute given to the Systems Science and Informatics Unit. We are thankful to the Freiburg Seizure Prediction Project for making their epileptic ECoG data available to the research community all over the world, which we eventually accessed and worked on. We would also like to thank the three anonymous referees whose suggestions resulted in improvements in the paper.